\documentclass[runningheads]{llncs}
\usepackage[T1]{fontenc}
\usepackage{graphicx}
\usepackage{adjustbox}
\usepackage{enumitem}
\usepackage{mathtools}
\usepackage{multicol, multirow}
\usepackage{booktabs}
\usepackage{amsmath}
\usepackage{amsfonts}

\usepackage{caption}
\usepackage{bbding}
\usepackage{enumitem}
\usepackage{subfloat}
\usepackage{graphicx}
\usepackage{xcolor}
\usepackage{subfig}
\usepackage[ruled,linesnumbered]{algorithm2e}
\usepackage{threeparttable}
\usepackage{bm}
\usepackage{array}

\usepackage[misc]{ifsym}

\newtheorem{myConstraint}{Constraint}
\begin{document}
\title{A Predict-Then-Optimize Customer Allocation Framework for Online Fund Recommendation\thanks{The first three authors contribute equally.}}
\titlerunning{A Predict-Then-Optimize Framework}
%

\author{Xing Tang\inst{1} \and
Yunpeng Weng\inst{1} \and
Fuyuan Lyu\inst{2} \and
Dugang Liu\inst{4} \and \\
Xiuqiang He{ \Letter}\inst{3}}
\authorrunning{Tang et al.}
%
\institute{FiT, Tencent, Shenzhen, China \\
School of Computer Science, McGill University, Montreal, Canada \\  
School of Big Data and Internet,Shenzhen Technology University, Shenzhen, China\\
Shenzhen University, Shenzhen, China \\
\email{\{shawntang,edwinweng\}@tencent.com} \\
\email{fuyuan.lyu@mail.mcgill.ca}; \email{\{dugang.ldg,he.xiuqiang\}@gmail.com} 
}
\maketitle              
\begin{abstract}

With the rapid growth of online investment platforms, funds can be distributed to individual customers online. The central issue is to match funds with potential customers under constraints.  Most mainstream platforms adopt the recommendation formulation to tackle the problem. However, the traditional recommendation regime has its inherent drawbacks when applying the fund-matching problem with multiple constraints. In this paper, we model the fund matching under the allocation formulation. We design \textbf{PTOFA}, a \textbf{P}redict-\textbf{T}hen-\textbf{O}ptimize \textbf{F}und \textbf{A}llocation framework. This data-driven framework consists of two stages, i.e., prediction and optimization, which aim to predict expected revenue based on customer behavior and optimize the impression allocation to achieve the maximum revenue under the necessary constraints, respectively. Extensive experiments on real-world datasets from an industrial online investment platform validate the effectiveness and efficiency of our solution. Additionally, the online A/B tests demonstrate PTOFA's effectiveness in the real-world fund recommendation scenario.

\end{abstract}

\section{Introduction}
\label{sec:introduction}

In recent years, selling funds via online platforms, such as Ant Fortune\footnote{https://www.antgroup.com/en/}, Tencent LiCaiTong\footnote{https://www.tencentwm.com} and Robinhood\footnote{https://robinhood.com}, has gained immense popularity due to their convenience of financial services. The core problem is achieving an optimal match between funds and customers, thereby fulfilling customers' interests and fund managers' requirements. Based on extensive investment records, these online financial platforms adopt data-driven approaches to improve their business and service quality as other online services~\cite{financial,value,fund,pimm}. 

\begin{figure*}[htbp]
	\centering
	\includegraphics[width=1.0\linewidth]{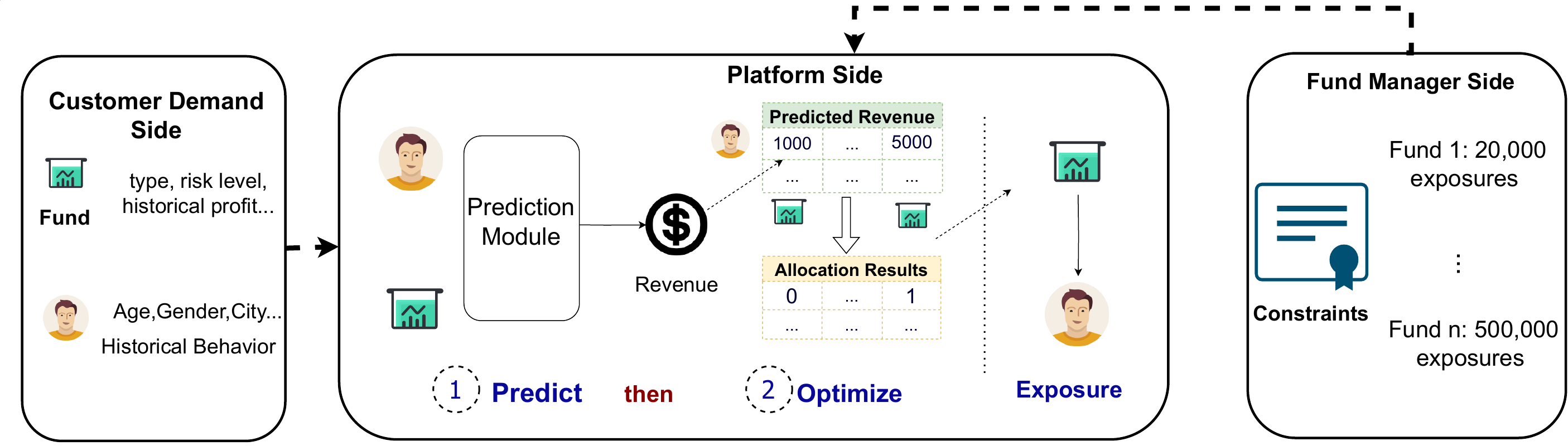}
	\caption{An illustration of fund matching process on an online investment platform. }
	\label{figure_process}
\end{figure*}

The matching problem between funds and customers on the platform is complex. As shown in Fig.~\ref{figure_process}, the platform must consider the requirements of three stakeholders: customers' interests, fund managers' requirements, and platforms' risk regulation.
To the end of customers, clicks and conversions don't fully reveal their interest in funds. Platforms should better predict customer revenue to gauge potential interest, as they tend to invest more in funds satisfying their needs.
To the end of fund managers, funds should be exposed to a certain amount of customers to guarantee their interest.
To the end of platforms, platforms must match risk levels between funds and customers, ensuring each fund only targets customers with higher risk tolerance. 
Under these constraints, platforms intend to maximize the revenue between customers and fund managers.

One possible solution is to formulate the problem as a top-\textit{k} recommendation~\cite{comb-k,fund_rec}.
The system will predict the relevance score measuring the customer-fund pair based on the customer's preference and the fund's properties. Top-$k$ funds will then be exposed to customers. Such a formulation aims to increase the conversion rate and the total number of transactions. However, we observe that this formulation may cause several problems: (i) \textit{Matthew effect}~\cite{matthew}. Exposure to customers is mainly occupied by a few top funds, leaving others few opportunities to be exposed. Meanwhile, some high-quality funds fail to be exposed to enough customers. Correspondingly, the majority of revenue also goes to these top funds. Such a phenomenon could harm the platform in the long run. (ii) \textit{Objectives mismatch}. The traditional recommender system aims to maximize the conversion of customers to items. We argue that this may result in a sub-optimal total revenue since they treat every conversion equally. However, in the fund matching problem, the revenue of the same customer may differ significantly on various funds. Both problems of the recommendation formulation motivate us to solve the fund matching problem by adopting an allocation~\cite{CAA,PTOC} formulation.

Instead of recommending suitable funds to each customer from a local perspective, we formulate the fund-matching problem as an allocation format, which can better encode constraints and maximize total revenue from a global perspective. We introduce the predict-then-optimize framework~\cite{pto,PTOC,weng2024expected} to allocate funds to each customer, as illustrated in Fig.~\ref{figure_process}.
However, we still have to face the following main challenges. \textit{How to predict expected revenue accurately?} Notice that the distribution of expected revenue is long-tailed because both the investment intention and available capital vary greatly regarding customer-fund pairs. Despite that mean squared error (MSE) has become the de facto loss for regression tasks, it will underestimate on imbalanced label~\cite{mse}. Moreover,  \textit{How to effectively allocate funds to customers under constraints?} In practical industrial applications, the delivery systems depend on manual operations~\cite{video} or reinforcement learning~\cite{PTOC}. As to manual operations, the allocation strategy determines the priority of $K$ products according to historical sales and operation specialists' experience. After that, it selects enough customers for each product to deliver by the order of priority. Therefore, the manual allocation strategy relies highly on human experience, which makes them sub-optimal. As to reinforcement learning, it is too complex to deploy in our scenario, which requires the model to converge to the optimal state quickly due to the low maintenance costs of funds. Also, the solution is hard to be easily extended to large-scale settings. 

In this paper, we design a two-stage Predict-Then-Optimize Fund Allocation (PTOFA) framework. During the first stage, PTOFA predicts the revenue brought by customer investments for each fund. In the second stage, PTOFA allocates the funds with a solver according to the predicted revenue in the first stage. Specifically, we first introduce the entire sample space for customer intention and expected revenue prediction to predict the revenue accurately. Then, we adopt a counterfactual multi-task learning module to predict the expected revenue in the entire sample space. Moreover, our framework takes all potential customer's predicted revenues for each fund as coefficients and employs an efficient optimization module to get the near-optimal allocation solution. In summary, the main contributions of our work are as follows:

\begin{itemize}


\item We highlight the inherent drawbacks of adopting recommendation formulation for the fund-matching problem. Instead, we propose to formulate it as the allocation problem. A predict-then-optimize fund allocation framework, namely PTOFA, is designed to match funds and customers on investment platforms under various constraints.

\item In PTOFA, we propose a counterfactual entire space multi-task module to predict the expected revenue. Note that it is the first attempt to predict the expected revenue of funds. We then design an efficient optimization algorithm based on the predicted revenue to get the solution.
	
\item Both offline and online experiments are conducted on LiCaiTong, one of the largest online investment platforms in China. Experimental results demonstrate the effectiveness and efficiency of our proposed PTOFA framework. 

\end{itemize}

\section{Formulation of Fund Allocation}
\label{sec:method_problem}
As we discussed in earlier sections, the core idea of fund allocation is to efficiently allocate various funds $f \in \mathbf{F}$ to different customers $u \in \mathbf{U}$ given customers' preferences and various constraints. We adopt a matrix $\mathbf{X} \in \{0, 1\}^{|\mathbf{U}| \times |\mathbf{F}|}$ to denote the allocation result, with $\mathbf{X}_{u,f} = 1$ indicating fund $f$ is allocated to customer $u$ and vice versa. The goal of the fund allocation problem is to maximize the total expected revenue income, and the objective function is formulated as follows:
\begin{equation}
\label{eq:loss_fund_alloc}
Obj =  \sum_{u\in \mathbf{U},f\in \mathbf{F}} \mathbf{X}_{u,f} \cdot \mathbb{E}_{u,f},
\end{equation}
where $\mathbb{E}_{u,f}$ is the revenue expectation of customer $u$ given fund $f$. The revenue expectation is predicted given the profile and historical record of customer $u$, as well as various properties of the fund $f$.

Such an allocation process is also subject to certain constraints. Similar to recommendation systems, only limited slots are available for exposing funds to customers. This can be defined as:
\begin{myConstraint}
\textbf{(Customer Top-K)} Each customer $u$ is only exposed to $K$ recommended funds:
$\sum_{f \in \mathbf{F}} \mathbf{X}_{u,f} = K, \ \forall u \in \mathbf{U}.$
\end{myConstraint}

Considering the long-term interest, the impression of customers should be distributed fairly among funds. This is defined as:
\begin{myConstraint}
\textbf{(Fund Exposure)} Each fund should be allocated to exactly $d_f$ customers, formulated as:
$\sum_{u \in \mathbf{U}} \mathbf{X}_{u,f} = d_f, \ \forall f \in \mathbf{F}.$
\end{myConstraint}
Finally, exposure to funds needs to be regulated due to the risky nature of the investment. Only funds with risk levels lower than customers' explicitly self-declared risk tolerance can be recommended to specific customers. This is defined as:
\begin{myConstraint}
\textbf{(Platform Risk)} $\mathbf{X}_{u,f}$ can only be 1 if the customer risk tolerance $t_u$ is no less than the risk level $r_f$.
$\mathbf{X}_{u,f} = 1 \ \rightarrow \ t_u \ge r_f, \forall u \in \mathbf{U}, \forall f \in \mathbf{F}.$
\end{myConstraint}


Hence, the goal of fund allocation is to maximize Eq. \ref{eq:loss_fund_alloc} under the above-mentioned constraints. However, both the $\mathbb{E}_{u,f}$ and $\mathbf{X}_{u,f}$ are not static during the process in Eq. \ref{eq:loss_fund_alloc}. An inaccurate prediction of the revenue expectation could mislead the fund allocation process. To solve this problem, we adopt a commonly used predict-then-optimize framework as a solution.
As its name explains, it consists of two tasks: prediction and optimization.

\textbf{Prediction.}
Accurately predicting the expected revenue requires the model to address two main issues: i) predict whether the customer will invest the delivered fund, and ii) if the conversion occurs, predict how much money the customer is willing to invest. Thus, we formulate the expected revenue as 
\begin{equation} \label{eq:ETV}
    \mathbb{E} = {\mathbf{P}(\text{C})}\times \mathbb{E}(R|C),
\end{equation}
where $C\in\{0,1\}$ denotes whether the customer is converted, $R$ denotes the revenue brought by the customer given the conversion, and $\mathbb{E}(R|C)$ is its expectation given that the customer converts.

\textbf{Optimization.}
After the prediction stage, for each customer $u$ and fund $f$, the $\mathbb{E}_{u,f}$ is regarded as the constant. The fund allocation can be viewed as a combinatorial optimization problem:
\begin{equation}
\label{eq:final_goal}
\begin{aligned}
    {\underset {\mathbf{X}}{\operatorname {arg\,max} }} &\sum_{u\in \mathbf{U},f\in \mathbf{F}} \mathbf{X}_{u,f} \cdot \mathbb{E}_{u,f}, \\
    \text{s.t.} & \ \text{Constraint 1, 2, \& 3}. 
\end{aligned}
\end{equation}


\section{Methodology}

\label{sec:method}

\begin{figure}[t] 
\centering    
\subfloat[Overall positive samples.]{
    \label{fig:distribution_a}
    \includegraphics[width=0.4\columnwidth]{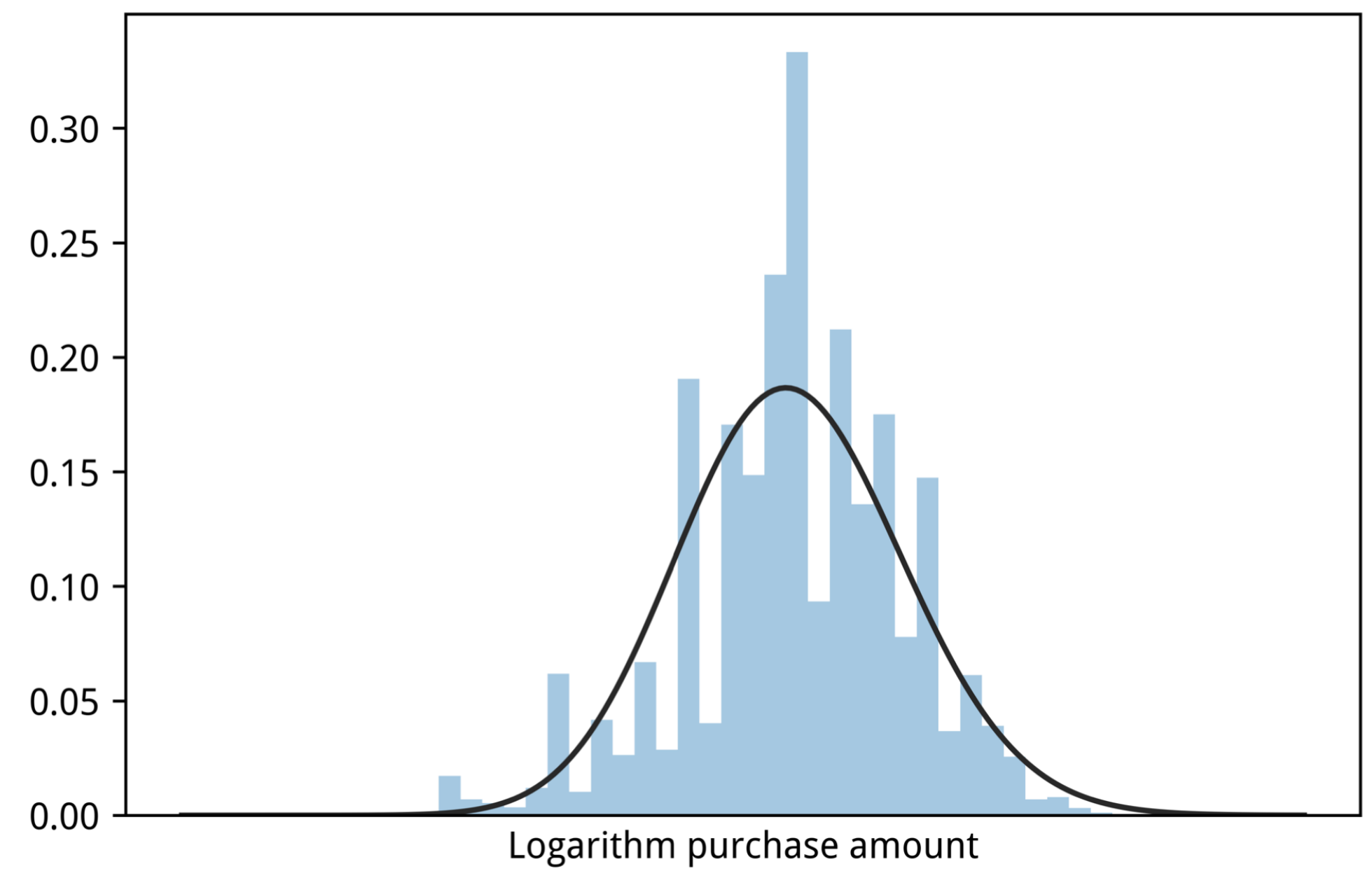}}
    \hspace{2pt}
\subfloat[Three users invest amount.] { 
    \label{fig:distribution_b}
    \includegraphics[width=0.4\columnwidth]{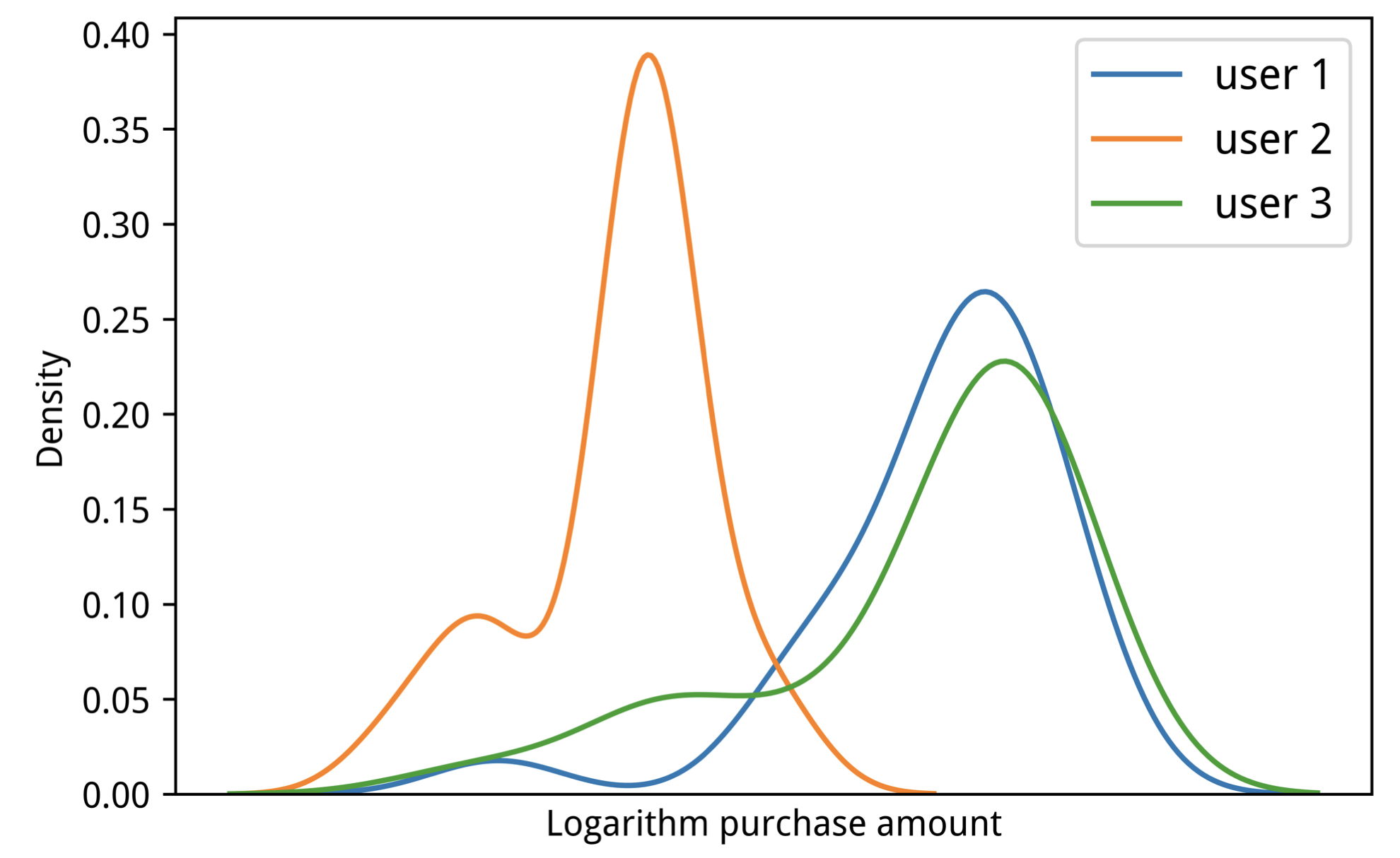}}    
    \vspace{-2pt}
\vspace{-8pt}
\caption{Distribution of logarithmic transaction value.}     
\label{fig:distribution}     

\end{figure}

\textbf{Revenue Prediction Over the Entire Space.} 
In the prediction stage, the prediction module needs to predict the conversion probability and the revenue brought by this conversion according to Eq.~\ref{eq:ETV} simultaneously. However, here are some unique challenges in the prediction task. First, imbalanced labels exist among converted customers. Specifically, customer revenue follows a long-tail distribution in real-world scenarios. Fig.~\ref{fig:distribution_a} shows the distribution of the logarithmic values of the revenue (purchase amount) brought by converted customers. Instead of a conventional regressor optimized by MSE loss, we adopt a parametric manner to model revenue distribution motivated by previous work~\cite{ziln,optdist}. 
Meanwhile, we also randomly present the historical investment amounts of three customers in Fig.~\ref{fig:distribution_b}, which follow log-normal distributions but with different parameters. Hence, the parameters of the distribution, i.e. the mean $\mu$ and the standard deviation $\sigma$, are estimated in the neural network with customer features and fund features. We can obtain the probability density function of converted samples' revenue $\mathbf{P}(R | C=1, \mu,\sigma) = {1}/({R\sqrt{2\pi}\sigma })\cdot\exp({-{(\log{R}- \mu)^2}/{2\sigma ^2}})$.

\begin{figure*} \centering    
    \subfloat[Training the regression task over positive sample space.]{
        \label{fig:previous_modeling}
        \includegraphics[width=0.465\columnwidth]{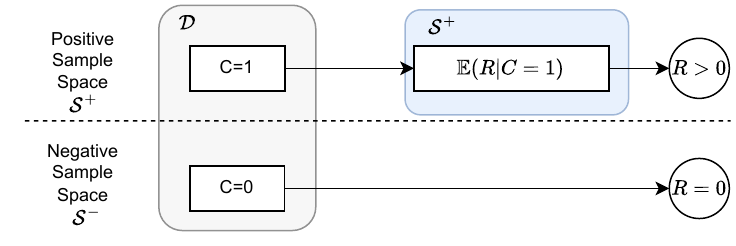}}  
        \hspace{15pt}
    \subfloat[Training the regression task over entire sample space.] { 
        \label{fig:do_modeling}
        \includegraphics[width=0.465\columnwidth]{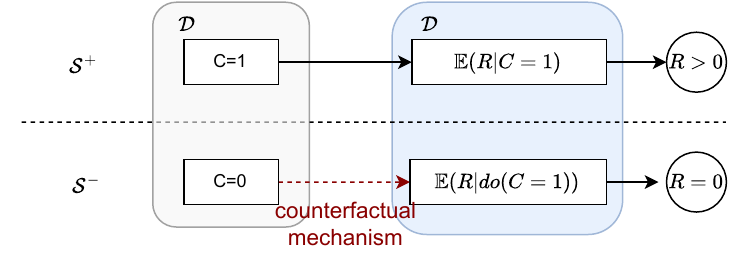}}    

    \caption{The existing solution and our idea.}  
    \label{fig:our_modeling}    

\end{figure*}

Second, previous works~\cite{ziln,yang2023feature} train multi-task learning model with conversion rate prediction task training over the entire sample space $\mathcal{D}$, while the regression task for revenue estimation is only trained over the positive sample space $\mathcal{S}^{+}$, as shown in Fig.~\ref{fig:previous_modeling}. This will lead to the \textbf{sample selection bias} \cite{esmm,escm2,dcmt} between training and inference, as the inference space is the entire sample space. Therefore, we utilize a counterfactual mechanism to debias the sample selection bias.  As shown in Fig.~\ref{fig:our_modeling}, we utilize the do-calculus\cite{dcmt} $do(C=1)$ to provide the counterfactual sample for regression task training over negative sample space $\mathcal{S}^{-}$. The $do$ indicates that we suppose the unconverted sample is converted and predicts its expected revenue.  Thus, we can train the prediction model in the entire space. To implement training in the counterfactual space, we need to set counterfactual labels for negative samples, i.e., the revenue $R$ if they convert. Since most customers have $R = 0$ after exposure and the counterfactual samples are all from negative sample space, we assume that the counterfactual revenue is a small value $\epsilon$.

Furthermore, we can formulate the probability of an observed positive sample with a customer-fund pair $(u, f)$:
\begin{equation}
\begin{aligned}
    \mathbf{P}( R | \vec{x}_u, \vec{x}_f)
    &=\mathbf{P}(C  = 1 | \vec{x}_u, \vec{x}_f)  \times \mathbf{P}(R| \vec{x}_u, \vec{x}_{f} , C  = 1) \\
    &= \frac{\text{P}_{c}}{R \cdot \sqrt{2\pi}\sigma  }\exp({-\frac{(\log(R) - \mu )^2}{2\sigma^2}}), 
    \label{positive} 
\end{aligned}
\end{equation}
where $P_c = \mathbf{P}(C = 1 | \vec{x}_u, \vec{x}_f)$ denotes the predicted conversion rate. The probability of an observed negative sample ($R=0$) can be formulated as:
\begin{equation}
\begin{aligned}
    \mathbf{P}(R = 0 | \vec{x}_u, \vec{x}_f) &=  1 - P_c + P_c \cdot \mathbf{P}( \epsilon | \vec{x}_u, \vec{x}_f, C=1) \\
    &= (1 - P_c) +  \frac{P_c}{\epsilon \cdot \sqrt{2\pi}\sigma }\exp({-\frac{(\log( \epsilon) - \mu)^2}{2\sigma ^2}}),	
    \label{negative} 
\end{aligned}
\end{equation}
We can thereby derive the following form of the entire space multi-task joint loss function from the negative log-likelihood: 
\begin{equation}
\begin{aligned}
    &\mathcal{L} = -\frac{1}{M}\Biggl(\sum_{(u,f) \in \mathbf{Y}_{\text{pos}}}\Bigl( \log P_c + \log \frac{1}{R \cdot \sqrt{2\pi} \sigma} - \frac{(\log(R) - \mu)^2}{2\sigma^2}\Bigr)  \\
 &+ \sum_{(u,f) \in \mathbf{Y}_{\text{neg}}}\Bigl( \log (1 - P_c + \frac{ P_c}{\epsilon\cdot \sqrt{2\pi}\sigma }\exp(-\frac{ (\log(\epsilon) -\mu )^2}{2\sigma ^2}))\Bigr)\Biggr).
\label{loss function} 
\end{aligned}
\end{equation}
Here $\mathbf{Y}_{\text{neg}} \equiv \{(u, f)|R = 0\}$ and $\mathbf{Y}_{\text{pos}} \equiv \{(u, f)|R > 0\}$ denote the negative and positive sets respectively. Throughout the proposed loss, the model has a joint probability modeling of the classification and regression tasks on the entire sample space. With the predicted $ P_c$, parameters of the log-normal distribution $\mu$ and $\sigma$, we can estimated the expected revenue for each $<u,f>$ pair: $\mathbb{E} = P_c \times (\exp(\mu + \sigma^2/2) )$.

\begin{algorithm}
    \caption{Optimization Algorithm $\textit{HA}$ of PTOFA}\label{algorithm}
    \KwIn{ fund set $\mathbf{F}$; matrix:\{$\mathbb{E}_{uf} | \forall u\in \mathbf{U}, \forall f\in \mathbf{F}$\}; customer risk tolerance: T = \{$t_u$|${u\in \mathbf{U}} $\};  fund risk level:R =\{$r_f$\}; exposure demend: D = \{$d_f$\}; K in top-K}
    \KwOut{$\mathbf{X}$}
    For all $<u,f>$ pair, if $r_f > t_u$, $ \mathbb{E}_{u,f} = -\infty$;
    
     $\forall f \in F$, $\alpha_f =  \sum_{u \in \mathbf{U}} \mathbb{E}_{uf}/{d_f}$ 
    
    \For{$u \in \mathbf{U}$}{
        Sort $\forall f \in \mathbf{F}$ in descending order based on $\mathbb{E}_{uf}$; 
        Compute the heuristic score $h_u$ according to Eq.~\ref{eq:hs}.
    }
    Sort $ \forall u \in \mathbf{U}$ in descending order based on $\{h_u\}$\;
    \While{$u \in |\mathbf{U}| $}{
        Select the top-K funds $f$ with highest $\mathbb{E}_{u,f}$ in terms of user $u$, $\mathbf{X}_{uf} = 1$ ,$\mathbf{U} = \mathbf{U} \backslash u$ , $d_f = d_f - 1$ \;
        \If{$d_f$ = 0}{
            $\forall u,  \mathbb{E}_{u,f} = -\infty , \mathbf{F} = \mathbf{F}\backslash f $ ; Execute from line 2\mbox{-}6;
        }
    }
\end{algorithm}

\textbf{Large Scale Fund Allocation Module}. 
The optimization stage solves Eq.~\ref{eq:final_goal} with $\mathbb{E}_{u,f}$  obtained in the prediction stage as constants. We proposed a specially designed heuristic algorithm (HA) in Algorithm \ref{algorithm} to tackle this problem under complex constraints. Specifically, a heuristic score is introduced to measure the potential loss of platform revenue.
For the customer $u$, we sort all funds in descending order of coefficient $\mathbb{E}$ according to the revenue matrix, getting the candidate fund list with ranking $\{f_1,f_2,...,f_{|F|}\}$ for the specific user and respectively calculate the predicted revenue difference between two adjacent funds. The final heuristic score is defined as follows:
\begin{equation}
\label{eq:hs}
    h_{u} =  {\sum_{j\in\{2,3,...,|F|\}}{\alpha_{f_{j-1}}}\times {(\mathbb{E}_{u,f_{j-1}} -\mathbb{E}_{u,f_j}})},
\end{equation}
where $\alpha_{f_i} =   \sum_{u \in \mathbf{U}} \mathbb{E}_{u,f_i} / d_{f_i}$ is introduced to estimate the consumption speed of the fund.
A large $\alpha_f$ indicates that the fund $f$ may reach the exposure limit sooner.
We set $\mathbb{E}_{u,f} =-\infty$ if the risk tolerance level of customer $u$ is lower than the risk level of fund $f$ to satisfy the \textit{Platform Risk Constraint}.  

Our heuristic algorithm (HA) in Algorithm \ref{algorithm} begins by allocating the allocated top-K funds $f$ for the user $u$ with the highest heuristic score and subsequently updates $d_{f}  = d_{f} - 1$ for these top-K funds. The allocation process then proceeds to the user with the second-highest score and so forth. Once a specific fund $f$ achieves allocation limited under the \textit{Fund Exposure Constraint}, we set $\mathbb{E}_{u,f} = -\infty$ for all remaining customers, ensuring that $f$ will not be allocated further. This process is repeated until all funds have been completely allocated.

\section{Experiments}

In this section, we conduct thorough experiments on the proposed method's effectiveness and efficiency. Because our method is a two-stage framework, predict-then-optimize, we compare some baselines on real-world offline datasets to verify that our prediction module is effective. Moreover, we compare different allocation optimizing strategies to show the superiority of the optimization module with the prediction module being determined. Finally, online experiments regarding the combination of various components are conducted.

\label{sec:experiment}
\begin{table}[!t]
\centering
\caption{Comparison of prediction models on Two real-world datasets.}
\label{offline:prediction}
\begin{tabular}{l|*{3}{>{\centering\arraybackslash}p{1.5cm}}||*{3}{>{\centering\arraybackslash}p{1.5cm}}}
\hline
 & \multicolumn{3}{c||}{LCT-A} & \multicolumn{3}{c}{LCT-B} \\ \hline
Model & AUC$\uparrow$ & MSE$\downarrow$ & MAE$\downarrow$ & AUC$\uparrow$ & MSE$\downarrow$ & MAE$\downarrow$ \\ \hline
ESMM & 0.8303  & -& - & 0.9270  & -& -  \\
DNN$\mbox{-}C$ & 0.8237& -& -& 0.9040& -& - \\
DNN$\mbox{-}R$ & - & 0.0386 & 0.0413 & -  & 0.4248 & 0.1369 \\
Shared$\mbox{-}$Bottom & 0.8120 & 0.0367 & 0.0323 & 0.9208 & 0.4265 & 0.1404   \\
MMoE  & 0.8205  & 0.0381 & 0.0392 & 0.9192 & 0.4236 & 0.1436  \\

ZILN &  0.8287  & 0.0386 & 0.0426& 0.9228 & 1.7885 & 0.5909  \\
MTL$\mbox{-}$MSE & 0.8276  & 0.0382 & 0.0353  & 0.9213  & 0.4262 & 0.1446  \\
$\mathbf{Ours}$ & \textbf{0.8327} & \textbf{0.0362} &\textbf{0.0227}  & \textbf{0.9313}&\textbf{ 0.4203} & \textbf{0.1326}  \\ \hline
\end{tabular}
\end{table}

\subsection{Offline Experiments}

\textbf{Prediction Module Evaluation}:  We use two real-world datasets from a large-scale platform to compare revenue prediction performance: \textbf{LCT-A} contains training(90\%) and validation (10\%) samples from historical promotional campaigns of three months, involving 8 promotional funds and 4.75 million users. The test set has 3.45 million users sampled from the promotional delivery of the subsequent week. \textbf{LCT-B} contains training (90\%) and validation (10\%) samples from recommendation scenario, involving 4 fund types and 4.3 million users. The testing set includes 0.69 million users.

We compare our approach with several widely used methods: Shared-Bottom ("Shared" for short)~\cite{mtl}, MMoE~\cite{ma2018modeling}, ESMM~\cite{esmm}, ZILN~\cite{ziln}.
Additionally, two DNN models are trained for conversion rate prediction and revenue estimation, denoted as \textit{DNN-C} and \textit{DNN-R} respectively.
The feature embedding sizes are set to 24 and the batch size to 512 for all cases. We conduct the grid search for learning rate in the range of \{0.01, 0.02, 0.04, 0.06\} and dropout in  \{0.1, 0.3, 0.5\} for each model, respectively. All models are optimized with Adam~\cite{kingma2014adam}.  AUC is used for conversion prediction while Mean Absolute Error (MAE) and Mean Squared Error (MSE) are used to evaluate revenue estimation~\cite{ltv_game,reliable}. 

Results reported in Table ~\ref{offline:prediction} illustrate that our model performs best in both conversion rate and revenue prediction tasks. 

\begin{figure}[t] \centering    
    \vspace{-5pt}
    \subfloat[The objective value.]{
        \label{obj}
        \includegraphics[width=0.45\columnwidth]{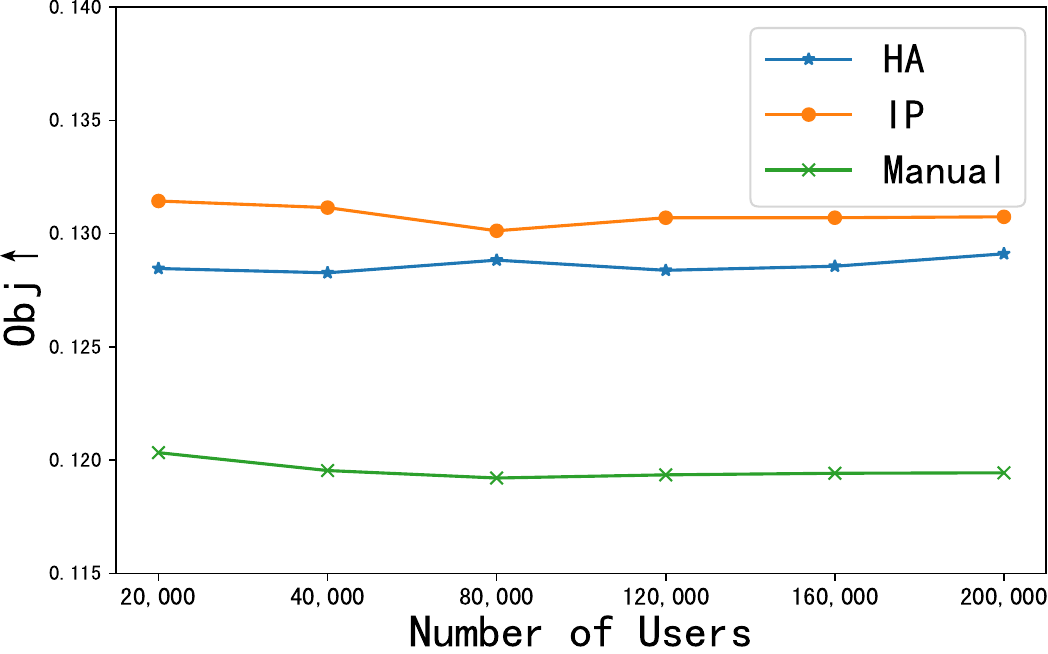}}     
    \subfloat[The time cost. ] { 
        \label{fig:time_cost}
        \includegraphics[width=0.45\columnwidth]{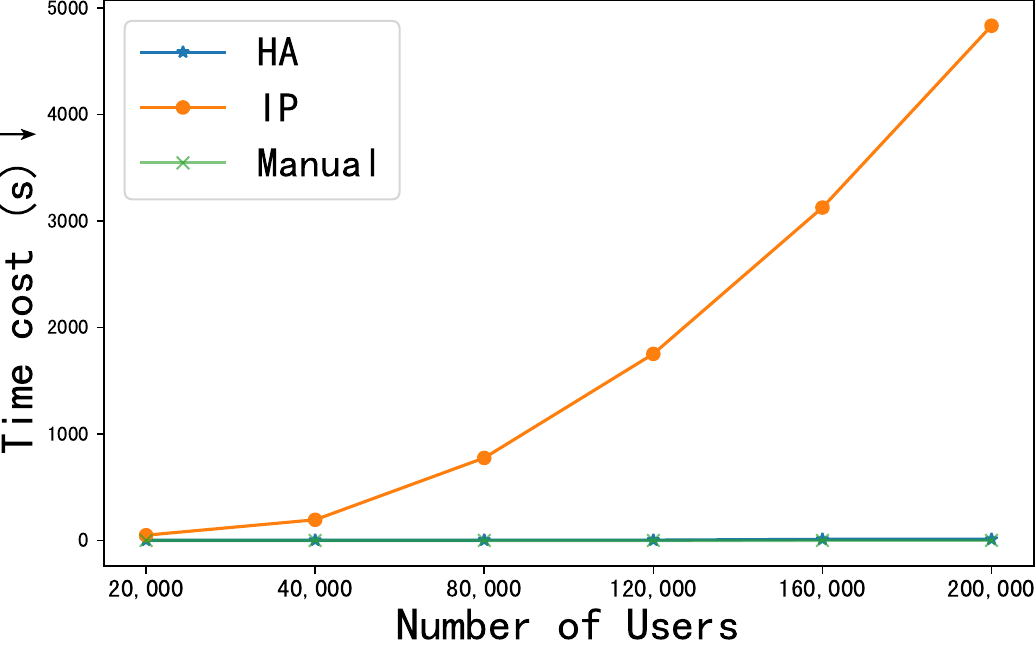}} 
    \vspace{-8pt}
    \caption{Comparison of different allocation strategies.}    

    \vspace{-15pt}
    \label{time_cost}     
\end{figure}

\textbf{Optimization Algorithm \textit{HA} Evaluation }:   

Following the prior work~\cite{comb-k}, we randomly sample datasets of varying user scales from the \textbf{LCT-A} dataset. For each sampled dataset with scale $S$, we linearly scale down the fund exposure constraint $d_f$ according to the sampling ratio and simulate the exposure allocation.  Two baselines are introduced as a comparison. (1)\textit{manual} allocation strategy. The priority of the funds is manually defined, given the funds' average historical conversion rate and the operation specialists' experience. Specifically, for the fund $f$ with the highest priority, we iteratively expose it to $d_f$ customers with the highest predicted revenue score until the allocation is completed. This is the baseline optimization algorithm, which is more similar to the traditional recommendation. (2)\textit{integer programming (IP)} solver, an off-the-shelf method for combinatorial optimization\cite{goptimize}.Fig. \ref{obj} shows that the IP achieves the best objective. Our proposed HA can obtain a similar performance as IP and can obtain significant improvement compared to manual allocation. For example, in the dataset with 200,000 users, HA achieves the \textbf{98.75\%} objective score of IP. Fig. \ref{fig:time_cost} shows the time cost of different strategies. The time cost of HA and Manual is close and continues to be significantly lower than IP in different data scales. In the dataset with 200,000 users, the speed-up ratio achieves \textbf{416}.  In our business, the allocation problem involves over 10 million users, and the solution time for traditional IP methods is unacceptable.

\subsection{Result from online A/B testings}

We have deployed our PTOFA on a large-scale online financial platform and conducted online A/B testing experiments on promotional campaigns. There are 10 million customers for each campaign, which is further divided into 5 million in the control and experimental groups, respectively. In each campaign, there are eight distinct candidate funds available for exposure, and each user will only be exposed to one of these funds (K=1).   For online A/B testing, we focus on two metrics: i) Conversion Per Mille Exposures (CPME) and ii) Revenue Per Mille Exposures (RPME), which are formulated as follows: 

\begin{equation}
	\begin{aligned}
		\mathbf{CPME} =\frac{ \mathbf{\# \ conversion}}{\mathbf{\#  \ exposures}} \times 1000, \quad
  \mathbf{RPME} =\frac{ \mathbf{\sum{revenue}}}{\mathbf{\#  \ exposures}} \times 1000.
		\label{metrics} 
	\end{aligned}
\end{equation}

\begin{table*}[!t]
\center
\caption{Online A/B testing results of different frameworks. }
\label{online}
\scalebox{1}{\begin{tabular}{ccccccc}
    \toprule
     Period & Predict &  Optimize & CPME & RPME& \%CPME Lift    & \%RPME Lift \\
    \midrule
    \multirow{2}{*}{ P1} & {ESMM}& Manual & 2.69 & 14,901 & - & -\\ 
    \cline{2-7} 
    &  ESMM & HA & 2.88 & 13,492 & 7.06\% & -9.46\%\\ 
    \hline
    
    \multirow{2}{*}{ P2} & {ESMM}& HA & 2.05 & 7,359 & - & -\\ 
    \cline{2-7} 
    &  Ours & HA & 2.55 & 10,048 & 24.39\% & 36.54\%\\ 
    \hline 
    \multirow{2}{*}{ P3} & {ESMM}& Manual & 3.39 & 16,608 & - & -\\ 
    \cline{2-7} 
    &  Ours & HA  & 4.29 & 25,558 & 26.54\% & 53.89\%\\ 
    \hline 
\end{tabular}}
\end{table*}

As the Table.~\ref{online} illustrated, our proposed PTOFA framework is superior to the current online baseline framework, which uses ESMM to predict the conversion rate after the fund $f$ exposed to user $u$ as $\mathbb{E}_{u,f}$ and employs the manual allocation strategy. Notice that the ESMM+Manual campaign recommends items in a traditional manner, which further indicates the significance of our proposed PTOFA framework.


\section{Related Work}
\label{sec:related_work}

\subsection{Generalized Recommendation and Fund Recommendation}

Generalized recommendations have been adopted in various online services, such as video distribution~\cite{video}, real estate~\cite{CAA}, and financial product~\cite{HTLNet,fund_rec}. Top-\textit{k} recommendation is a widely researched topic, where the goal is to recommend a list of items to users that users may be interested in~\cite{comb-k}. It is able to enlarge the engagement of users, and is more practical in accords to real-world recommendation scenarios. Starting from matrix factorization techniques~\cite{BPR} to learn latent representation of users and items, increasing research focus on deep neural network-based methods. NeuMF~\cite{NCF} and NGCF~\cite{NGCF} utilize multi-layer perception and graph neural networks to boost performance. This also motivates research to deploy top-\textit{k} recommendations on fund distribution. Graphical deep collaborative filtering has been used in fund recommendation~\cite{fund_rec,fund_rec2,fund}. However, we argue that generalized recommendation leads to some problems for fund matching on online investment platforms, motivating a shift from recommendation to allocation. 

\subsection{Predict-then-Optimize Problem}

Predict-then-Optimize(PTO) is summarized as predicting unknown parameters based on historical data and generating decisions by solving the corresponding optimization problem via the predicted parameters. SPO~\cite{pto} is a general framework for solving this problem by proposing an SPO loss to measure the error in predicting the cost vector of the optimization problem. Many applications involve PTO problems, e.g. couriers allocation~\cite{PTOC}, portfolio investment~\cite{portfolio}. PTOC~\cite{PTOC} is proposed to solve the couriers allocation for emergency last-mile logistics. It predicts user demand with a variational graph GRU encoder and optimizes the courier task allocation problem with multi-agent graph reinforcement learning. As far as we know, we are the first to adopt PTO in fund allocation.


\section{Conclusion}
\label{sec:conclusion}

In this paper, we propose a Predict-Then-Optimize Fund Allocation framework that transforms the fund-matching problem from recommendation to allocation. The PTOFA consists of two stages. In the prediction stage, we propose a revenue prediction module that deals with how to predict revenue over the entire space. In the optimization stage, we design an efficient heuristic algorithm to allocate the funds under the necessary constraints, i.e., \textit{customer top-k}, \textit{fund exposure}, and \textit{platform risk}. Extensive online and offline experiments on a real investment platform confirm the effectiveness and efficiency of our proposed framework.

\bibliographystyle{splncs04}
\bibliography{reference.bib}

\end{document}